\documentclass[sigconf,natbib=false]{acmart}
\pdfoutput=1
\AtBeginDocument{%
  }

\setcopyright{acmcopyright}
\copyrightyear{2018}
\acmYear{2018}
\acmDOI{XXXXXXX.XXXXXXX}

\RequirePackage[
  datamodel=acmdatamodel,
  style=acmnumeric,
  ]{biblatex}

\usepackage[utf8]{inputenc} 
\usepackage[T1]{fontenc}

\usepackage[frozencache,cachedir=.]{minted}
\definecolor{bg}{rgb}{1,1,1} 
\setminted{
    fontfamily=zlmtt,    
    fontsize=\footnotesize, 
    baselinestretch=1.0,
    breaklines, 
    autogobble,
    bgcolor=bg,
    frame=lines,
    framesep=2mm,
}

\usepackage[ruled,linesnumbered, noend]{algorithm2e}

\makeatletter
\let\@float@c@listing\@caption
\makeatother

\renewcommand\footnotetextcopyrightpermission[1]{}
\begin{document}

\title{PyHGL: A Python-based Hardware Generation Language Framework}

\author{Jintao Sun}
\affiliation{%
  \institution{Zhejiang University}
  \country{China}}
\email{jintaos2@zju.edu.cn}

\author{Zeke Wang}
\affiliation{%
  \institution{Zhejiang University}
  \country{China}
}

\author{Tao Lu}
\affiliation{%
  \institution{Zhejiang University}
  \country{China}
}

\author{Wenzhi Chen}
\affiliation{%
  \institution{Zhejiang University}
  \country{China}
}

\renewcommand{\shortauthors}{Trovato et al.}

\begin{abstract}
Hardware generation languages (HGLs) increase hardware design productivity by creating parameterized modules and test benches. Unfortunately, existing tools are not widely adopted due to several demerits, including limited support for asynchronous circuits and unknown states, lack of concise and efficient language features, and low integration of simulation and verification functions. This paper introduces PyHGL, an open-source Python framework that aims to provide a simple and unified environment for hardware generation, simulation, and verification. PyHGL language is a syntactical superset of Python, which greatly reduces the lines of code (LOC) and improves productivity by providing unique features such as dynamic typing, vectorized operations, and automatic port deduction. In addition, PyHGL integrates an event-driven simulator that simulates the asynchronous behaviors of digital circuits using three-state logic. We also propose an algorithm that eliminates the calculation and transmission overhead of unknown state propagation for binary stimuli. The results suggest that PyHGL code is up to 6.1$\times$ denser than traditional RTL and generates high-quality synthesizable RTL code. Moreover, the optimized simulator achieves 2.9$\times$ speed up and matches the performance of a commonly used open-source logic simulator. 
\end{abstract}



\keywords{Hardware Description Language, Hardware Generation, Python, Hardware Simulation}


\maketitle

\section{Introduction}
In modern hardware design, non-traditional architectures, such as reconfigurable and heterogeneous architectures, are increasingly being adopted to improve performance under strict power and area constraints. However, traditional hardware description languages (HDLs) like Verilog \cite{verilog}, and VHDL \cite{vhdl} cannot satisfy the demand for fast design space exploration and rapid prototyping because of their limited semantics and design patterns. Two major approaches were proposed to improve hardware design productivity. First, High-Level-Synthesis (HLS) is a technique that automatically compiles software-oriented descriptions into low-level hardware implementations. However, the HLS code optimization is usually time-consuming \cite{hls} and may not result in optimal designs \cite{hlslimit}. Second, Hardware Generation Languages (HGLs) construct hardware from configurable and extensible modules on register-transfer abstraction \cite{rethink}. This approach not only facilitates code reuse but also speeds up verification by starting from verifying small-scale designs.

\setlength{\textfloatsep}{2mm}% Remove \textfloatsep

\begin{figure}[t]
    \centering
    \includegraphics[width=.9\linewidth]{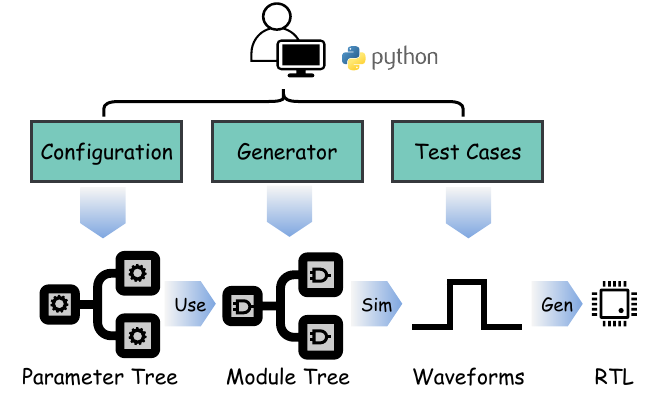}
    \vspace{-4mm}
    \caption{Overview of PyHGL's flow. Parameterization, generation, debugging, and verification are all integrated in the Python environment.}
    \Description{Parameterization, generation, debugging, and verification are all integrated in the Python environment.}
    \label{flow}
    % \vspace{-8mm}
\end{figure} 

%  \cite{golden}
HGLs are typically embedded in meta-programmed host languages in order to inherit the meta-programming features \cite{golden}. For example, Chisel\cite{chisel} based on Scala, PHDL\cite{phdl} based on Python, and ShakeFlow\cite{shakeflow} based on Rust. Python is an appropriate host language for three reasons. Firstly, Python's dynamic typing provides flexibility for both design and usage of the HGL. Secondly, as one of the most popular languages, Python has rich libraries and a large user base, which decreases the learning cost. Thirdly, Python's readable and minimalist coding style (known as ``Pythonic'') can also benefit hardware generation. Unfortunately, previous Python-based HGLs have revealed practical challenges in both readability and completeness. For example, MyHDL \cite{myhdl} and Mamba \cite{mamba} translate the Abstract Syntax Trees (ASTs) of Python functions to hardware implementations, which results in intermixed software-hardware semantics that reduces readability. PyRTL \cite{pyrtl} and Magma \cite{magma} are another two Python-based HGLs building hardware from a core set of primitives. However, they only support a single clock domain, restricting their application to asynchronous circuits. In addition, none of these Python-based HGLs nor Chisel \cite{chisel} support unknown states that plays an important role in hardware verification and optimization. 

This paper introduces PyHGL, a Python-based hardware generation language framework for modeling, simulation, and verification. First, PyHGL introduces necessary structures and operators for hardware design by extending Python's syntax and generating a modified parser using Pegen \cite{pegen}, the official parser generator of Python. The additional syntax is fully compatible with Python code. Second, PyHGL uses three-state (0, 1, and X) logic and implements a flexible intermediate representation (IR) capable of modeling asynchronous circuits. To improve the productivity of universal hardware design, PyHGL provides Pythonic language features such as dynamic typing, vectorized operations, and automatic port deduction. The complete PyHGL workflow is shown in Fig.~\ref{flow}.

Additionally, PyHGL integrates an event-driven simulator that performs behavior-level simulation directly on the intermediate representation. The integrated simulator avoids the semantic gap caused by using RTL simulators and provides Python's powerful language features for testing and debugging. Coroutine-based simulation tasks and concurrent assertions are supported for rapid verification. Furthermore, we introduce an algorithm that eliminates the massive computation and transmission overhead of X-propagation when the circuit contains only a few unknown states. 

The main contributions of this paper are as follows: (a) We introduce PyHGL, a full-featured hardware generation language embedded in Python that supports modular, configurable, and object-oriented design patterns (Section \ref{overview}). Language features such as dynamic typing, vectorized operations, and automatic port deduction are introduced to improve productivity. Minor semantics, such as dynamic assignments, latches, and tri-state wires, are well supported. (b) We implement a three-state and event-driven simulator that performs delay-accurate simulation for asynchronous digital circuits (Section \ref{simulator}). Verification features such as simulation tasks and assertions are supported. An algorithm that selectively executes X-propagation is introduced to increase the simulation performance. (c) We evaluate PyHGL framework (Section \ref{evaluation}). PyHGL language is both complete and efficient, found to be 6.1$\times$ denser than Verilog, 1.7$\times$ denser than Chisel, and 1.4$\times$ denser than Magma. Moreover, PyHGL simulator matches the performance of Icarus Verilog, a commonly used open-source logic simulator.

\section{PyHGL Language}\label{overview} 

PyHGL language provides both gate-level and register-transfer-level abstractions. Unlike Verilog, which uses an \texttt{always} block with an explicit sensitive list to describe the behavior of a variable, PyHGL uses pre-defined gates and only provides non-blocking assignments. Such an approach is more clear from the perspective of hardware designers, while the register-transfer abstraction remains available by providing conditional assignments.

\subsection{PyHGL IR}

PyHGL introduces a flexible intermediate representation that models digital circuits as a direct graph of \texttt{Gate} and \texttt{SignalData} nodes. A \texttt{Gate} can have an arbitrary number of inputs, sensitive inputs, and outputs. A \texttt{SignalData} can drive multiple gates but cannot be mult-driven. \texttt{Writer} is the edge from gate to data, and \texttt{Reader} is the edge from data to gate. The instance of \texttt{Reader} or \texttt{Writer} also contains a \texttt{SignalType} which defines the type information of the signal. Since the signal data and type are decoupled, type castings are simple. PyHGL provides a complete set of \texttt{Gate}s and \texttt{SignalType}s for constructing and manipulating complex asynchronous circuits.

% \begin{listing}[t]
% \begin{minted}[xleftmargin=6pt,linenos,frame=none,numbersep=2pt]{python}
% uint8_t  = UInt[8]
% sint32_t = SInt[32]
% vector_t = 4 ** uint8_t 
% mem_t    = MemArray[1024,8]
% struct_t = Struct(
%     a = vector_t  @ 0,  # field 'a' starts from 0
%     b = sint32_t  @ 0,  # field 'b' starts from 0 
%     c = UInt[32]  @ 32, # field 'c' starts from 32
% )                       # struct_t is 64-bit
% x = struct_t({
%     'a': [1,2,3,4], 
%     'c': Logic('xxxx1100'),
% })                      # signal x is 64'hxc04030201
% y = UInt[64](x)         # type casting
% \end{minted}
% \vspace{-8mm}
% \caption{PyHGL Signal Types and Type Casting}
% \label{hgltype}
% \vspace{-4mm}
% \end{listing}

\subsection{Literals and Signals}

The basic literal type of PyHGL is \texttt{Logic}, representing a three-state (0, 1, and X) value with arbitrary length. While the X state does not exist in real digital circuits, it is used in simulation to represent undefined behaviors, such as dividing by zero or bit selection out of range. Besides, the intentional assignment of X to a signal prompts the synthesizer to perform further optimizations. PyHGL does not have a high-impedance (Z) state, which is only used in tri-state wires and is considered less significant in digital circuits. Instead, we represent an undriven tri-state wire using the X state. Python literals, like strings and integers, can be automatically transformed into the \texttt{Logic} type. PyHGL also includes a \texttt{BitPat} literal type representing three states: 0, 1, and don't care, which is only used in comparison. PyHGL has a rich type system. \texttt{UInt[w]} defines an unsigned integer type of fixed width \texttt{w}. Composite types \texttt{Vector} and \texttt{Struct} are similar to packed arrays and structs in Verilog, which contain information of bit fields. And \texttt{MemArray} defines multidimensional unpacked arrays that usually model memories.

\subsection{Functions and Gates}\label{functions}

Gates are generated when calling functions and operators on signals. To address the issue that Python lacks sufficient operators for logic design, PyHGL overrides existing operators and introduces some new operators. For example, the \texttt{!}, \texttt{\&\&} and \texttt{||} operators indicate logic operations NOT, AND, and OR, respectively, while \texttt{<==} and \verb|<=>| indicate directional assignments and nondirectional connections. PyHGL provides a rich set of functions, including bitwise operations, reductions, comparisons, and arithmetic operations. All functions and operators are dynamically dispatched via a map from argument types to registered functions. As a result, operations on different type of signals are well-defined and customizable during the building stage. 

\begin{listing}[t]
\begin{minted}[xleftmargin=6pt,linenos,frame=none,numbersep=2pt]{python}
# io is an array of shape (2,2)
io = Bundle(
    x = Bundle(
        a = UInt[8](0),
        b = SInt[8](0)),
    y = Bundle(
        a = UInt[8](0),
        b = SInt[8](0)),
) 
# out is an array of two registers
with ClockDomain(clock=(clk,1), reset=(rst_n,0)):
    out = Reg(UInt[8].zeros(2))
# identical operations  
out <== io[:,'a'] + io[:,'b'] 
out <== Array([io.x.a + io.x.b, io.y.a + io.y.b])
\end{minted}
\vspace{-8mm}
\caption{PyHGL Array, Clock Domain, Vectorized Operation and Assignment Example.}
\label{array}
% \vspace{-4mm}
\end{listing}

PyHGL implements a container type \texttt{Array} for vectorized operations. \texttt{Array} is tree-like and can store both elements and names. \texttt{Array} elements can be accessed using advanced slicing methods. Most of PyHGL's functions and operators are default vectorized. Specifically, iterable arguments are converted into \texttt{Array}s, and the function is mapped on each dimension of \texttt{Array} arguments. Vectorized operations significantly reduce the usage of explicit loops, decrease lines of code, and increase code readability. Listing~\ref{array} demonstrates how to construct named arrays using \texttt{Bundle} function (lines 2-9) and apply vectorized operations (lines 14-15). \texttt{UInt} and \texttt{SInt} signals from \texttt{io} are added up respectively, and result in an array of two signals. The result is assigned to \texttt{out} through the operator \texttt{<==}. PyHGL supports commonly used netlists in the Verilog standard. Netlists are gates that support connection semantics. In PyHGL, combinational netlist \texttt{Wire}s are implicitly generated in assignments, while other netlists including \texttt{Reg}, \texttt{Latch}, and tri-state wire \texttt{Wtri} should be explicitly declared. For convenience, default clock and reset signals are provided, and arbitrary clock and reset are supported through \texttt{ClockDomain}. In Listing~\ref{array} on lines 11-12, it defines two registers that are triggered at the positive edge of signal \texttt{clk} with a negative asynchronous reset signal \texttt{rst\_n}.

\subsection{Assignments and Modules}

\begin{listing}[t]
\begin{minted}[xleftmargin=6pt,linenos,frame=none,numbersep=2pt]{python} 
@module VendingMachine:
    nickel = UInt(0) @ Input    # 5 cent coin
    dime   = UInt(0) @ Input    # 10 cent coin
    valid  = UInt(0) @ Output   # valid when >= 20 cents
    switch s:=Reg(EnumOnehot()): 
        once 'sIdle':
            when nickel: s <== 's5'
            when dime  : s <== 's10' 
        once 's5':
            when nickel: s <== 's10'
            when dime  : s <== 's15'
        once 's10':
            when nickel: s <== 's15'
            when dime  : s <== 'sOk' 
        once 's15': 
            when nickel: s <== 'sOk'
            when dime  : s <== 'sOk'
        once 'sOk':
            s, valid <== 'sIdle', 1
\end{minted}
\vspace{-8mm}
\caption{Vending Machine Example.}
\label{vending}
% \vspace{-4mm}
\end{listing}

PyHGL introduces a new operator \texttt{<==} for hardware assignments. Assignments are directional, non-blocking, and have priority. Conditional assignments avoid the use of explicit multiplexers. PyHGL uses the \texttt{when} construct to represent branches and the \texttt{switch} construct to match signal values. Both constructs generate a ``condition'' gate that supports X-propagation. PyHGL supports the ``reverse case'' statement and ``unique case'' flag for further optimization. An example is ``\texttt{switch (1, Flag.unique): ...}'', which provides a  non-priority assumption that helps the synthesizer generate more optimized hardware. Dynamic assignments are statements that assign to part of a signal indexed by another signal, which is rarely supported by other HGLs. An example of this is ``\texttt{a[idx::8] <== b}'', which updates 8 bits of the signal \texttt{a} starting from \texttt{idx}. In PyHGL, a signal can be assigned inside or outside the module, and all assignments will be collected in order.

Finite state machines (FSM) are widely used in digital circuit design. PyHGL provides a dynamic enumerated type that simplifies the declaration of state machines and reduces the efforts for modifying state encodings. Unlike basic signal types with a fixed bit length, an enumerated type has a variable length. It accepts a string literal as a new state and implicitly maps it to a binary, one-hot, or gray encoded value. As shown in Listing~\ref{vending}, a vending machine is fully defined within one \texttt{switch} block, where \texttt{EnumOnehot()} (line 5) generates a state register \texttt{s} with a one-hot-encoded enumerated type. The enumerated type finally records five states: `sIdle', `s5', `s10', `s15', and `sOK'. 

\begin{listing}[t]
\begin{minted}[xleftmargin=6pt,linenos,frame=none,numbersep=2pt]{python}
from pyhgl.logic import *
@conf Config:
    @conf RippleCarry:
        w = 32
    @conf KoggeStone:
        w = 64
AdderIO = lambda w: Bundle(
    x   = UInt[w  ](0) @ Input,
    y   = UInt[w  ](0) @ Input,
    out = UInt[w+1](0) @ Output)
@module FullAdder: 
    a, b, cin = UInt([0,0,0])
    s    = a ^ b ^ cin 
    cout = a & b | (a ^ b) & cin 
@module RippleCarry:
    io = AdderIO(conf.p.w) 
    adders = Array(FullAdder() for _ in range(conf.p.w))
    adders[:,'a'  ] <== io.x.split()
    adders[:,'b'  ] <== io.y.split()
    adders[:,'cin'] <== 0, *adders[:-1,'cout']
    io.out <== Cat(*adders[:,'s'], adders[-1,'cout']) 
@module KoggeStone:
    io = AdderIO(conf.p.w) 
    P_odd = io.x ^ io.y
    P = P_odd.split()
    G = (io.x & io.y).split()
    dist = 1 
    while dist < conf.p.w:
        for i in reversed(range(dist,conf.p.w)): 
            G[i] = G[i] | (P[i] & G[i-dist])
            if i >= dist * 2:
                P[i] = P[i] & P[i-dist]
        dist *= 2 
    io.out <== Cat(0, *G) ^ P_odd
@task tb(self, dut, N): 
    for _ in range(N):
        x, y = setr(dut.io[['x','y']]) 
        yield self.clock_n() 
        self.AssertEq(getv(dut.io.out), x + y)
with Session(Config()) as sess:
    adder1, adder2 = RippleCarry(), KoggeStone()
    sess.track(adder1, adder2)  
    sess.join(tb(adder1, 100), tb(adder2, 200))
    sess.dumpVCD('Adders.vcd') 
    sess.dumpVerilog('Adders.sv') 
\end{minted}
\vspace{-8mm}
\caption{Example of modeling and simulating a Ripple Carry adder and a Kogge Stone adder.}
\label{adder}
% \vspace{-4mm}
\end{listing}

PyHGL modules are defined using the \texttt{@module} structure, the syntactic sugar for decorated functions. Signals within modules are recorded during module instantiation and can be accessed outside of the module. Listing~\ref{adder} provides an example that showcases the design and verification of a Ripple Carry adder and a Kogge Stone adder. In PyHGL, signals from different modules can be freely connected, and module ports can be automatically inferred when elaborating RTL code. Therefore, explicit port declarations are unnecessary for simple modules. For instance, the \texttt{FullAdder} is implemented in only three lines of code (lines 11-14) without any explicit port. PyHGL offers two parameterization methods: passing parameters as module arguments and using the parameter tree. A parameter tree is top-down inherited and has the same structure as the module tree. The \texttt{@conf} statement is used to define a configuration function that is executed when the module name matches. For instance, the \texttt{Config} function (lines 2-6) will insert \texttt{w=32} to the Ripple Carry adder and \texttt{w=64} to the Kogge Stone adder. Parameters for the current module can be accessed through \texttt{conf.p} (lines 16, 23, 28, 29).

\section{Simulation and Verification}\label{simulator} 

PyHGL uses an integrated simulator for two reasons. First, third-party simulators such as Verilator \cite{verilator} and Icarus Verilog \cite{icarus} require hand-written RTL test benches, which leads to a semantic gap in the verification stage. Second, useful verification features like assertions can be easily implemented within Python. The PyHGL simulator is event-driven and supports accurate gate delays. An event queue of signal update events on discrete time slots is maintained. For each time slot, signals are updated and triggered gates are executed.

\subsection{Coroutine-based Simulation Tasks}\label{task}

PyHGL simulation tasks are non-synthesizable sequential processes for verification purposes. They are essentially Python generator functions (functions that contain \texttt{yield} keyword), a subset of Python coroutines that can be paused and resumed asynchronously. The PyHGL simulator accepts coroutine-based tasks and executes them concurrently with the gate simulation. Like Verilog \texttt{task} blocks, PyHGL simulation tasks support variable triggers such as time delay and signal edges. A simulation task can also call and wait for other tasks to finish using the \texttt{join} function. In Listing~\ref{adder} on lines 35-39, a task \texttt{tb} that verifies the functionality of an adder is defined using the \texttt{@task} structure. The function \texttt{setr} sets random values to signals (line 37), and the function \texttt{getv} returns the current values of signals (line 39). On line 43, the task \texttt{tb} is used to verify both the Ripple Carry adder and Kogge Stone adder.

\begin{table}[t]
\small
\begin{tabular}{p{0.14\textwidth}p{0.295\textwidth}}
\toprule
\textbf{Pattern} &  \textbf{Explanation} \\ 
\midrule
\texttt{n}  & Wait for \texttt{n} clock edges. \\ 
\hline \texttt{[m,n]}  & Wait for \texttt{m} to \texttt{n} clock edges. \\ 
\hline \texttt{edge(clk)}  & Wait until next edge of \texttt{clk}. \\ 
\hline \texttt{until(signal, v)} & Wait until value of \texttt{signal} is \texttt{v}. \\ 
\hline \texttt{\{name:signal\}}   & Cache signal values.\\
\hline \texttt{patt**n} & Repeat \texttt{patt} for n times. \\ 
\hline \texttt{patt**[m,n]} & Repeat \texttt{patt} for \texttt{m} to \texttt{n} times. \\ 
\hline \texttt{!patt} & Success if \texttt{patt} does not match. \\ 
\hline \texttt{patt1 \&\& patt2} & Success if both patterns match. \\ 
\hline \verb|patt1 >>> patt2| & Success if both patterns match. \\ 
\hline \texttt{patt1 || patt2} & Success if either pattern matches. \\ 
\hline \texttt{patt1 |-> patt2} & Success if \texttt{patt1} fails or both patterns success. \\ 
\hline \texttt{signal} &  Success if signal value is non-zero. \\ 
\hline \texttt{==, !=, >, <} &  Comparisons \\ 
\bottomrule
\end{tabular}
\vspace{1mm}
\caption{PyHGL Assertion Patterns}
\label{patterns}
\vspace{-8mm}
\end{table}

\begin{listing}[t]
\begin{minted}[xleftmargin=6pt,linenos,frame=none,numbersep=2pt]{python}
with Assertion(trigger=(clk,1), disable=(rst,0)) as a:
    # en should fell once rose
    p1 = a.rose(en) |-> 1 >>> a.fell(en)
    # write twice and read once on same address
    p2 = a.rose(en) && w && {'addr': addr} |-> (
        2 >>> a.rose(en) &&  w && addr==a.get('addr'),
        2 >>> a.rose(en) && !w && addr==a.get('addr') 
    )
    a.check(p1, p2)
\end{minted}
\vspace{-8mm}
\caption{Example of PyHGL Assertions}
\label{assert}
% \vspace{-4mm}
\end{listing}

\subsection{Assertions}

PyHGL supports concurrent assertions with semantics similar to SystemVerilog Assertions (SVA). Assertions provide a lightweight and powerful way to verify hardware at multiple stages of the design flow. PyHGL assertions are triggered and evaluated at each clock edge during the simulation. New operators \verb|>>>| and \texttt{|->} are introduced for sequence and implication semantics, respectively. PyHGL assertions are achieved based on Python coroutines and the dynamic dispatch mechanism discussed above. Like regular expressions that match string patterns, PyHGL assertions match signal values and hardware behaviors over time. There are two kinds of patterns: assertions on signals that either succeed or fail, and commands or actions that always succeed. Table~\ref{patterns} lists some supported patterns and commands. Listing~\ref{assert} provides some examples of PyHGL assertions. The \texttt{Assertion} context (line 1) not only sets up the trigger and the disable signals but also switches the dispatcher to change the operator functionalities. Property \texttt{p1} (line 3) asserts that signal \texttt{en} should fall one clock cycle after it rises. Property \texttt{p2} asserts that it should write twice, followed by a read to the same address. These two properties are asserted every time the trigger arrives (line 9).

\subsection{Accelerate Logic Simulation}

\begin{figure}[t]
    \centering
    \includegraphics[width=.94\linewidth]{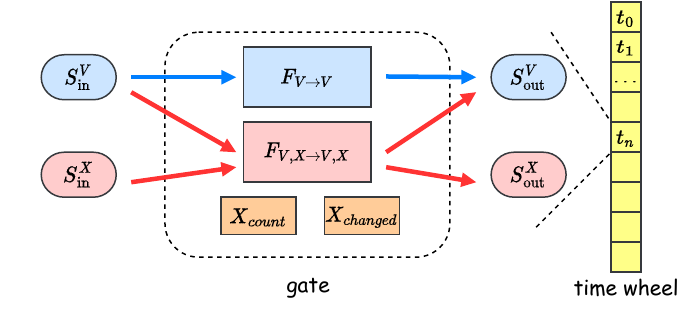}
    \vspace{-4mm}
    \caption{PyHGL Simulation Functions of Each Gate.}
    \Description{simulation}
    \label{gate}
    % \vspace{-4mm}
\end{figure}

Although the X state plays an important role in both verification and optimization, the calculation and transmission overhead is massive. Two bits are required to represent the three-state logic: \texttt{00} as zero, \texttt{10} as one, and both \texttt{01} and \texttt{11} as the unknown state (X). In an event-driven simulator, X-propagation doubles the signal updates and increases the gate execution time because triggered gates must compute both binary and X outputs based on the three-state inputs. 

In actual designs, X states are rarely produced. We also found that most combinational gates (excluding dividers and part selects) have binary outputs for binary inputs. Therefore, we can define two simulation functions for each gate. As shown in Fig.~\ref{gate}, $S^V_{\text{in}}$ is the binary bits of gate input, $S^X_{\text{in}}$ is the unknown bits of gate input, $S^V_{\text{out}}$  and $S^X_{\text{out}}$ are output bits. $F_{V,X \to V,X}$ performs a complete calculation of three-state logic, and $F_{V \to V}$ performs a simpler binary calculation and only updates $S^V_{\text{out}}$. During simulation, $F_{V \to V}$ is executed if there is no change on unknown inputs $S^X_{\text{in}}$ and all of them are zero; otherwise, $F_{V,X \to V,X}$ is executed. Two extra variables $X_{count}$ and $X_{changed}$ are maintained for each gate. $X_{count}$ stores the number of non-zero X inputs, and $X_{changed}$ indicates whether any X input has changed at the current time slot. Algorithm~\ref{algorithm} describes the detailed signal update and gate execution flow for each time slot. Triggered gates are collected in $G_t$. The $waiting$ flag is used to avoid duplicated gate execution. $X_{count}$ and $X_{changed}$ are updated during the signal update stage. In the gate execution stage, a three-state simulation is only necessary either $X_{count}$ or $X_{changed}$ is non-zero.

\begin{algorithm}[t]
\caption{Optimized three-state logic simulation}\label{algorithm}
\KwData{current time $t$, current signal update events $E_t$}
$G_t \leftarrow [\ ]$\;
\ForEach{$(value, target)\in E_t$}{
    \If{$target$ \textnormal{is unknown bits} and $target \ne value$}{
        \textnormal{Update} $target$\;
        \textnormal{Evaluate} $\Delta X_{count}$\; 
        \ForEach{$g \in target.fanouts$}{
            $g.X_{count} \leftarrow g.X_{count} + \Delta X_{count}$\; 
            $g.X_{changed} \leftarrow 1$\; 
            $g.waiting \leftarrow 1$\; 
            \textnormal{Append} $g$ to $G_t$
        }
    }
    \If{$target$ \textnormal{is binary bits} and $target \ne value$}{
        \textnormal{Update} $target$\;
        \ForEach{$g \in target.fanouts$}{
            $g.waiting \leftarrow 1$\;
            \textnormal{Append} $g$ to $G_t$
        }
    }
}
\ForEach{$g \in G_t$}{
    \If{$g.waiting$}{
        \If{$g.X_{count}$ or $g.X_{changed}$}{
            \textnormal{Execute three-state simulation on} $g$\; 
            $g.X_{changed} \leftarrow 0$\;
        } \Else{
            \textnormal{Execute binary simulation on} $g$\; 
        }
        $g.waiting \leftarrow 0$\;
    }
}
$t \leftarrow t + 1$\;
\vspace{-0.1mm}
\end{algorithm}

Edge-triggered gates such as registers should always perform three-state simulations since some of their inputs are not in the sensitive list, so the simulator cannot maintain a valid $X_{count}$. Fortunately, there is only transmission overhead for registers, and combinational gates are usually much more than registers in real designs. Though some special gates, such as dividers and dynamic part selects, may generate X states for binary inputs, they are rarely used and therefore have little impact on the overall performance. In an event-driven behavior simulator, the overhead of Algorithm~\ref{algorithm} is negligible: only one extra comparison to decide the target type in the signal update stage, and two extra comparisons on $X_{count}$ and $X_{changed}$ to select the simulation function in the gate execution stage. Overall, the algorithm can effectively increase the performance of a three-state simulator when there are few unknown states in the circuit.

\section{Evaluation}\label{evaluation} 

PyHGL framework aims to provide a unified agile hardware design environment in Python. Section~\ref{comp} shows the completeness of the PyHGL framework. Section~\ref{dev} compares the development efforts of different languages using lines of code (LOC), a widely used metric in the software community. In Section~\ref{simperf}, we evaluate the effectivity of Algorithm~\ref{algorithm} by testing the relative performance of PyHGL simulators using different simulation strategies on a 128-bit Wallace tree multiplier. All tests are evaluated in the same platform running CPython-3.11 and Icarus-12.0.

\subsection{PyHGL Completeness}\label{comp} 

To evaluate the completeness of the PyHGL framework, we implemented both synchronous and asynchronous designs, including shift registers using NAND gates, asynchronous FIFO, adders, multipliers, AES ciphers, and simple RISCV cores. We simulated these designs in the PyHGL simulator and Icarus Verilog \cite{icarus}, a full-featured Verilog simulator that also supports accurate delay and unknown state. The PyHGL simulator used coroutine-based simulation tasks as described in Section~\ref{task}, while Icarus Verilog used hand-written RTL test benches. Our results show that PyHGL generates waveforms with asynchronous glitches identical to Icarus Verilog. Table~\ref{resource} compares the FPGA resource utilizations between PyHGL-generated and hand-written Verilog codes for AES\cite{aes} and SRV32\cite{srv32} designs. In AES implementation, PyHGL uses fewer Lookup Tables (LUTs) and Flip-Flops (FFs) than the hand-written Verilog. While in SRV32 implementation, PyHGL uses fewer LUTs, but slightly more FFs. The results suggest that PyHGL generates high-quality synthesizable RTL code.

\begin{table}[t]
\small
\centering
\begin{tabular}{|c|c|c|c|c|}
\hline                  & \multicolumn{2}{|c|}{AES} & \multicolumn{2}{|c|}{SRV32} \\
\hline                  & LUT     & FF     & LUT      & FF      \\
\hline \textbf{PyHGL}   & 2739      & 2745        & 7191       & 1983         \\
\hline \textbf{Verilog} & 2999      & 2748        & 7378       & 1952   \\ 
\hline
\end{tabular}
\vspace{2mm}
\caption{LUT and FF utilizations of AES and SRV32 implementations on a specified FPGA} \label{resource}
\vspace{-8mm}
\end{table}

\subsection{Development Productivity}\label{dev}

\begin{figure}[t]
    \centering
    \includegraphics[width=.92\linewidth]{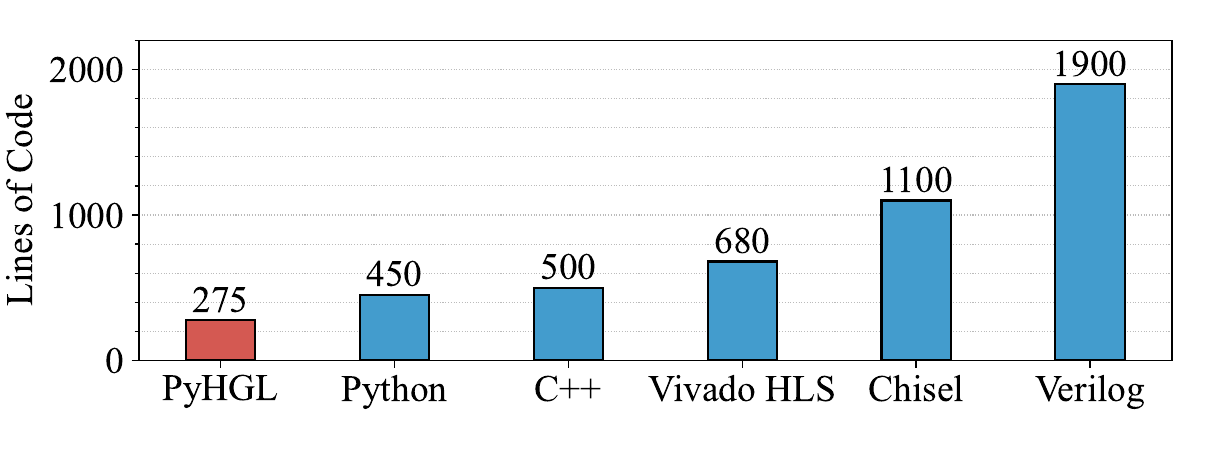}
    \vspace{-4mm}
    \caption{LOC comparisons among PyHGL AES and five AES implementations provided by Agile-AES \cite{guo2022agile}.  }
    \Description{loc}
    \label{AESloc}
    % \vspace{-4mm}
\end{figure}

We rewrote Usselmann's \cite{aes} Verilog implementation of AES encipher and decipher in PyHGL. Both implementations have the same architecture, the same ports, and output the same waveforms. The Verilog implementation is 1690 LOC long, while the PyHGL version is only 275 LOC, which is 6.1$\times$ denser. The Verilog test bench is 284 LOC, while the PyHGL test bench is 67 LOC, which is 4.2$\times$ denser. The PyHGL implementation mainly benefits from its compact vectorized operations described in Section~\ref{functions}, which are extensively used in the AES algorithm. Fig.~\ref{AESloc} also compares the PyHGL AES with other five designs provided by Agile-AES \cite{guo2022agile}, including hardware implementations in Chisel, Vivado HLS, Verilog, and software implementations in C++ and Python. Despite differing in implementation details, they are all complete and well-tested, making the comparison meaningful. The PyHGL implementation is 6.9$\times$ denser than Verilog, 4.0$\times$ denser than Chisel, 2.5$\times$ denser than Vivado HLS, 1.8$\times$ denser than C++, and 1.6$\times$ denser than Python.

In PyHGL, we implemented 13 examples from the Chisel tutorial \cite{chiseltutorial}, such as a Game of Life, simple ALUs, and a Router. The PyHGL code is 227 lines, which is 1.7$\times$ denser than the Chisel implementation (396 LOC) and 1.4$\times$ denser than the Magma \cite{magma} version (338 LOC). Compared with these two HGLs, PyHGL's advantage comes from vectorized operations and more signal types. In the SRV32 \cite{srv32} example, a 3-stage pipelined RISC-V processor implemented in 1093 lines of Verilog code (excluding one header file for common constants), the PyHGL version is 517 LOC long, which is 2.1$\times$ denser. Predefined multiplexers, simplified port connections, and vectorized assignments are used in the PyHGL implementation to decrease development efforts.

\begin{figure}[t]
    \centering
    \includegraphics[width=.9\linewidth]{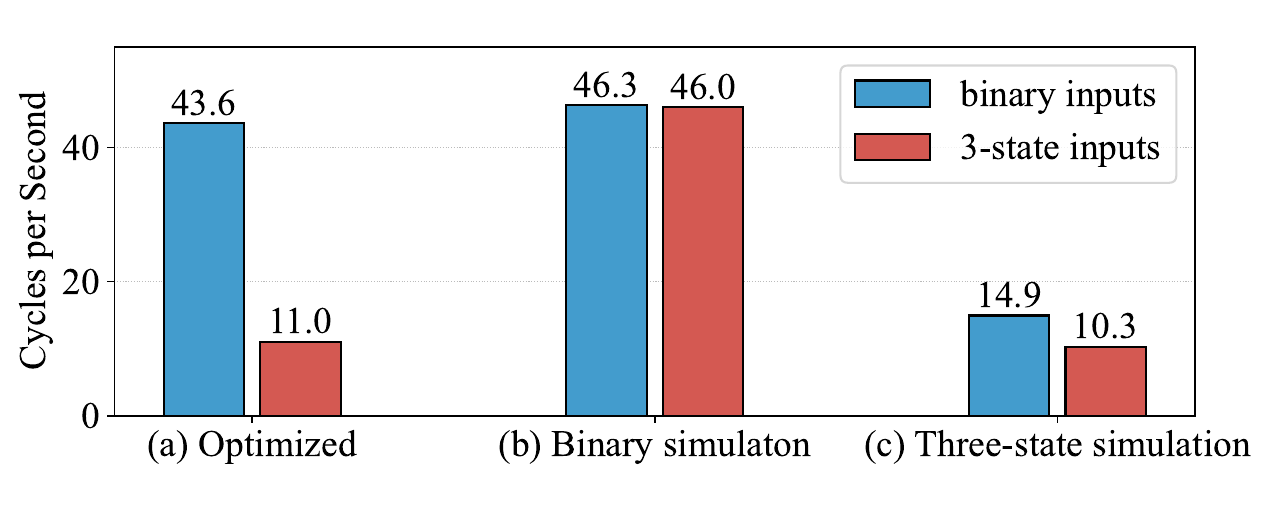}
    \vspace{-6mm}
    \caption{Quantitative comparison of simulation speed using different strategies on a 128-bit Wallace tree multiplier.}
    \Description{sim}
    \label{sim}
    \vspace{-3mm}
\end{figure}

\subsection{Simulator Performance}\label{simperf}

We developed three versions of the PyHGL simulator: (a) optimized using Algorithm~\ref{algorithm}; (b) always performs binary simulation; and (c) always performs three-state simulation. Fig.~\ref{sim} compares these three simulators' performance (Cycles per Second, CPS) on a 128-bit Wallace tree multiplier using different stimuli. For binary stimuli, the optimized simulator performs similarly to the binary simulator and is 2.9$\times$ faster than the three-state simulator. For three-state stimuli, the binary simulator cannot generate the correct waveform, while the optimized simulator and the three-state simulator have similar performance. The results indicate no significant overhead of Algorithm~\ref{algorithm} on both binary and three-state simulation.

\begin{figure}[t]
    \centering
    \includegraphics[width=.9\linewidth]{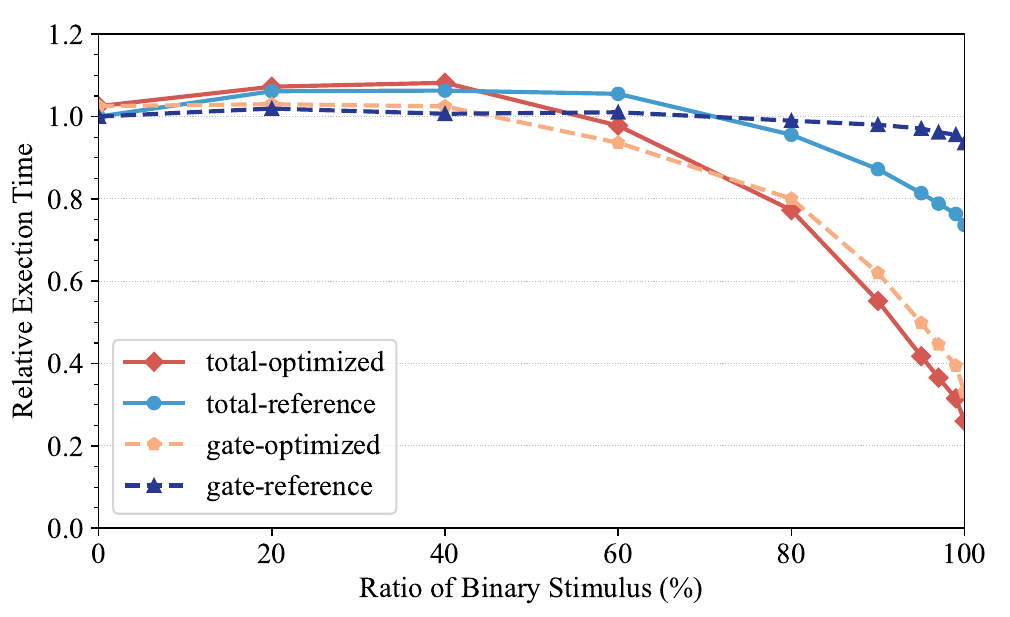}
    \vspace{-4mm}
    \caption{Execution Time vs. Ratio of Binary stimuli on the Wallace tree multiplier.}
    \Description{simref}
    \label{simref}
    \vspace{-2mm}
\end{figure} 

\begin{table}[t]
\small
\centering
\begin{tabular}{|c|c|c|c|}
\hline                 &  Adder & Divider & AES128/256 \\
\hline \textbf{PyHGL Simulator}  & 31.6s  & 18.1s    & 1.74s      \\ 
\hline \textbf{Icarus Verilog}  & 32.5s & 34.2s & 0.62s      \\
\hline
\end{tabular}
\vspace{2mm}
\caption{Simulation time of PyHGL on a 512-bit adder, a 512-bit divider and the AES cipher compared with Icarus Verilog.} \label{perf2}
\vspace{-4mm}
\end{table}

Fig.~\ref{simref} compares the execution time of the optimized simulator and the three-state reference simulator concerning the ratio of binary stimuli. The solid lines compare the total execution time, while the dotted lines compare the average execution time per gate. As the ratio of binary stimuli increases, the reference simulator has almost unchanged gate execution time, while the optimized simulator costs significantly less execution time. The total simulation time of both simulators increases first because it is also affected by the gate activation rate, which first increases and then decreases. When all stimuli are binary, the optimized simulator is about 2.8$\times$ faster than the reference simulator. The result suggests that the optimized simulator can be significantly accelerated when the circuit has a high ratio of binary states.  

Table~\ref{perf2} shows the performance of the optimized PyHGL simulator in terms of simulation time when compared to Icarus Verilog, which simulates generated RTL codes. All three tests use binary stimuli. The PyHGL simulator performs similarly to Icarus Verilog on the 512-bit Ripple Carry Adder. It is significantly faster on the 512-bit iterative divider, while Icarus Verilog performs better in the AES example. The CPython interpreter has a high overhead in maintaining the event queue, and this proportion decreases when the gate computation time increases. Therefore, the PyHGL simulator performs relatively better when the signals have bigger bit-lengths. For example, the PyHGL simulator has 899k CPS on a 4096-bit divider, which is 15$\times$ faster than Icarus Verilog's 59k CPS.

\section{Conclusion}\label{conclusion}

This paper introduced PyHGL, a state-of-the-art Python-based HGL framework for streamlined hardware design, simulation, and verification. PyHGL introduces well-designed syntax and implements a flexible intermediate representation, enabling it to concisely describe complex and asynchronous circuits. PyHGL provides Pythonic language features and powerful verification tools to achieve the agile design pattern. Additionally, PyHGL's event-driven simulator is accelerated by selectively performing unknown state propagation. The whole PyHGL framework is open source.

\printbibliography

\end{document}